\documentclass[%
 reprint,
 amsmath,amssymb,
 aps,
]{revtex4-2}

\usepackage{graphicx}
\usepackage{dcolumn}
\usepackage{bm}

\begin{document}

\preprint{APS/123-QED}

\title{Autocorrelation Measurement of Attosecond Pulses Based on Two-Photon Double Ionization}
\author{Fei Li$^{1,2}$}
\author{Kun Zhao$^{2,3,\ast}$}
\author{Bing-Bing Wang$^{2,4}$}
\author{Xin-Kui He$^{2,3,4}$}
\author{Zhi-Yi Wei$^{2,3,4,}$}
 \email{Corresponding authors: zhaokun@iphy.ac.cn; zywei@iphy.ac.cn}
\affiliation{$^{1}$Research Center for Advanced Optics and Photoelectronic, Department of Physics, College of Science, Shantou University, Shantou, Guangdong 515063, China\\
$^{2}$Beijing National Laboratory for Condensed Matter Physics, Institute of Physics, Chinese Academy of Sciences, Beijing 100190, China\\
$^{3}$Songshan Lake Materials Laboratory, Dongguan 523808, China\\
$^{4}$University of Chinese Academy of Sciences, Beijing 100049, China}



\date{\today}

\begin{abstract}

Autocorrelation measurement is theoretically demonstrated to characterize attosecond pulses by studying the two-photon double ionization (TPDI) process. An interferometric autocorrelation curve is presented in the change of TPDI probability with the time delay between two identical attosecond pulses, and its full width at half maximum (FWHM) $\tau_{e}$ has a relationship $\tau_{e}=1.77\tau+15$ with the FWHM $\tau$ of the attosecond pulse. The curve is also decoded to obtain the center frequency and FWHM of the attosecond pulse by fitting. In addition, the required peak intensity of the attosecond pulse is estimated to be on the order of $10^{16}\,\rm{Wcm^{-2}}$ in autocorrelation experiments. The findings pave the way for autocorrelation measurement of intense isolated attosecond pulses.

\end{abstract}

\maketitle


\section{\label{sec:level1}Introduction}

Since the isolated attosecond pulse was first generated in the laboratory\cite{hentschel2001attosecond}, many new technologies\cite{sola2006controlling,jullien2008ionization,mashiko2008double,wang2023fast,goulielmakis2008single} have been proposed to generate attosecond pulses with shorter pulse width\cite{goulielmakis2008single,sansone2006isolated,zhao2012tailoring,li201753,gaumnitz2017streaking}. At the same time,  investigations of electron dynamics on extremely short time scales in atoms, molecules and condensed matter have been carried out in virtue of attosecond pulses\cite{schultze2010delay,ossiander2017attosecond,hassan2016optical,calegari2014ultrafast,cavalieri2007attosecond,zhang2009attosecond,dienstbier2023tracing,loriot2024attosecond,severino2024attosecond}. Obviously, obtaining accurate information of attosecond pulses, such as pulse duration, frequency, peak intensity, chirp, is crucial for a wide range of applications. However, characterization of complete parameters of attosecond pulses is challenging. So far, the most common characterization method of attosecond pulses is the attosecond streak camera\cite{hentschel2001attosecond,itatani2002attosecond}, which detects the photoelectrons excited by an attosecond pulse and modulated by an infrared laser pulse and is a type of measurement of cross correlation between the attosecond extreme ultraviolet  and femtosecod infrared pulses. Another scheme which is all-optical is to measure the spectra of the attosecond pulse generated by the fundamental laser pulse and perturbed by a second-order harmonic or weak laser pulse\cite{kim2013manipulation,yang2020all}.  These characterization methods contain the principle of cross correlation measurement.

Besides cross correlation measurement, autocorrelation measurement stands as another vital method in the realm of signal processing. Gliserin et al. suggested that by introducing an intensity asymmetry into a nonlinear interferometric autocorrelation, it was possible to preserve certain spectral phase information within the autocorrelation signal, thereby enabling the complete reconstruction of the original electric field\cite{gliserin2022complete}. Previously, autocorrelation measurement has been used to characterize femtosecond pulses\cite{paye1993measurement,ranka1997autocorrelation,mitzner2009direct,osaka2022hard},  attosecond pulses trains\cite{nabekawa2006interferometric,nikolopoulos2005second,papadogiannis2003feasibility} and a specific harmonic order \cite{nabekawa2005production,nakajima2002use}. Therefore autocorrelation measurement of photoelectron signal excited by attosecond pulses may be a route to characterize isolated attosecond pulses as well, if the attosecond pulses are intense enough to trigger a two-photon process. In this paper we demonstrate theoretically that autocorrelation measurement based on the two-photon double ionization (TPDI) can be used to characterize intense isolated attosecond pulses.

\section{Theory}
The interaction between two time-delayed attosecond pulses and a helium atom is studied by solving the time-dependent Schr\"odinger equation (TDSE). The specific process to solve this TDSE has been described in detail elsewhere \cite{Fei2019Undestanding,Fei2020Universality}, so only the main idea is introduced here. In the velocity gauge and the electric dipole approximation, the TDSE reads (atomic units are used throughout, unless otherwise stated)
\begin{equation}
\label{equation1}
\begin{aligned}
i\frac{\partial}{\partial t}\Phi(\textbf{r}_{1},\textbf{r}_{2},t)=[H_{0}+\textbf{A}(t)\cdot(\textbf{p}_{1}+\textbf{p}_{2})]\Phi(\textbf{r}_{1},\textbf{r}_{2},t),
\end{aligned}
\end{equation}
where $H_{0}$ is field-free Hamiltonian of the helium atom, $\textbf{A}(t)$ is the vector potential of the attosecond pulses. The vector potential is expressed as
\begin{equation}
\label{equation2}
\begin{aligned}
\textbf{A}(t)&=A_{1}e^{-(2\ln 2)t^{2}/\tau^{2}}\sin(\omega t+\phi_{1})\hat{\textbf{e}}_{z}+\\
                &\hspace{5mm}A_{2}e^{-(2\ln 2)(t-t_{d})^{2}/\tau^{2}}\sin[\omega (t-t_{d})+\phi_{2}]\hat{\textbf{e}}_{z},
\end{aligned}
\end{equation}
where $A_{i}$ and  $\phi_{i}$ ($i=1,\,2$) are the amplitude and the carrier-envelope phase (CEP), $\tau$ is the full width at half maximum (FWHM), $\omega$ is the central frequency, $t_{d}$ is the time delay of the two attosecond pulses, and $\hat{\textbf{e}}_{z}$ is the unit vector of the polarization direction. The two-electron time-dependent wave function $\Phi(\textbf{r}_{1},\textbf{r}_{2},t)$ can be expanded in terms of eigenfunctions of $H_{0}$, then substituting the expanded formula into Eq.~(\ref{equation1}), a set of  coupled differential equations are presented, which can be solved by the Adams method\cite{Shampine1975Computer}. Once the time-dependent wave function $\Phi(\textbf{r}_{1},\textbf{r}_{2},t)$ is determined,  the energy distribution of two ionized  electrons at the time $t_{f}$ is written as
\begin{equation}
\label{equation3}
\begin{aligned}
P(E_{1},E_{2})&=\iint |\langle\varphi_{\textbf{k}_{1},\textbf{k}_{2}}(\textbf{r}_{1},\textbf{r}_{2})|\Phi(\textbf{r}_{1},\textbf{r}_{2},t_{f})\rangle|^{2}\times\\
&\hspace{5mm}k_{1}k_{2}d\hat{\textbf{k}}_{1}d\hat{\textbf{k}}_{2},
\end{aligned}
\end{equation}
where $\varphi_{\textbf{k}_{1},\textbf{k}_{2}}(\textbf{r}_{1},\textbf{r}_{2})$ is the uncorrelated double continuum state\cite{Fei2019Undestanding,Fei2020Universality}, $E_{1}$ and $E_{2}$ are the energies of two ionized electrons. Therefore, the probability of double ionization is $P_{t}=\iint P(E_{1},E_{2})dE_{1}dE_{2}$.

According to the time-dependent perturbation theory (TDPT), the energy distribution of the two ionized electrons is  \cite{WeiChao2014Double,Stefanska2012Two,Palacios2009Two} 
\begin{equation}
\label{equation4}
\begin{aligned}
P(E_{1},E_{2})&=\frac{1}{2}\Big(\frac{c}{4\pi^{2}}\Big)^{2}\Bigg|\sqrt{\frac{\sigma^{\rm{He}}(E_{1})}{\omega_{\alpha i}}}\sqrt{\frac{\sigma^{\rm{He^{+}}}(E_{2})}{\omega_{f\alpha}}}K(E_{\alpha})\\
&\hspace{5mm}+\sqrt{\frac{\sigma^{\rm{He}}(E_{2})}{\omega_{\beta i}}}\sqrt{\frac{\sigma^{\rm{He^{+}}}(E_{1})}{\omega_{f\beta}}}K(E_{\beta})\Bigg|^{2},
\end{aligned}
\end{equation}
where $c$ is the speed of light, and $\sigma^{\rm{He}}(E)$ and $\sigma^{\rm{He^{+}}}(E)$ are the one-photon single ionization (OPSI) cross sections of He atom and $\rm{He^{+}}$ ion, respectively. $\omega_{\alpha i}=E_{\alpha}-E_{i}$, $\omega_{f\alpha}=E_{f}-E_{\alpha}$,  $\omega_{\beta i}=E_{\beta}-E_{i}$, $\omega_{f\beta}=E_{f}-E_{\beta}$,  $E_{i}=-I_{p}=-2.903\,\rm{a.u.}$ and $E_{f}=E_{1}+E_{2}$ are the energies of the initial and final states, respectively. $E_{\alpha}=E_{1}-I_{p2}$ and $E_{\beta}=E_{2}-I_{p2}$ are the energies of the intermediate states,  $I_{p1}=0.903\,\rm{a.u.}$ and $I_{p2}=2\,\rm{a.u.}$ are the first and second ionization potentials, and $I_{p}=I_{p1}+I_{p2}$ is the ionization potential for ionizing two electrons. The function $K(E_{\alpha})$ is written as
\begin{equation}
\label{equation5}
\begin{aligned}
K(E_{\alpha})=\int_{-\infty}^{\infty}d\tau_{1}F(\tau_{1})e^{i\omega_{f\alpha}\tau_{1}}\int_{-\infty}^{\tau_{1}}d\tau_{2}F(\tau_{2})e^{i\omega_{\alpha i}\tau_{2}},
\end{aligned}
\end{equation}
where $F(t)=-\partial A(t)/\partial t$ is the electric field of the attosecond pulses. The $K(E_{\beta})$ is obtained by replacing the subscript $\alpha$ with $\beta$ in Eq.~(\ref{equation5}). It is clear that the validity of TPDI restricts the frequency  in the range $\omega>I_{p2}$. The TDSE is accurate but time-consuming, while the TDPT is fast but underestimates electron correlation and the number of intermediate states in the case of ultrashort pulses\cite{WeiChao2015Virtual}. Therefore, in this work, the probability of TPDI is mainly obtained by the TDPT, and the result of the TDSE serves as a standard reference. 

\section{Results And Discussion}

\begin{figure}[!tb]
\includegraphics[width=75mm]{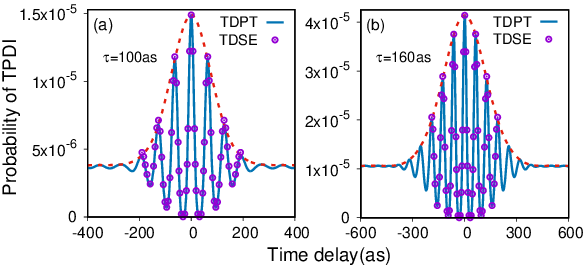}
\caption{\label{figure1} The IAC curve of photoelectrons generated by TPDI during the interaction between a helium atom and two time-delayed attosecond pulses. The blue solid line and purple circles represent the results of TDPT and TDSE, respectively, and the red dashed line shows the upper-envelope. The two attosecond pulses are identical. The peak intensity, center frequency and CEP of the attosecond pulses are $1\times10^{14}\,\rm{Wcm^{-2}}$, 2.4 a.u. and 0, respectively. The FWHM of the attosecond pulses are: (a) 100 as; (b) 160 as. }
\end{figure}

The feasibility of autocorrelation measurement of attosecond pulse based on the TPDI is indicated by the fact that the TPDI is a second order nonlinear process. Furthermore, based on our previous work\cite{Fei2019Undestanding}, the energy distribution of the two emitted  electrons in the TPDI appears as either a single or two elliptical peak. By fitting the contour curve that corresponds to the half maximum probability density, the lengths of the semi-major $a$ and minor axes $b$ are obtained. In particular, the relationship $b=\sqrt{2}/\tau$ exists in the energy distribution of two emitted electrons in the TPDI, which means the information of the attosecond pulse is really carried by the two ionized electrons. The ionization probability of TPDI is selected as the measured quantity in the autocorrelation measurement of attosencond pulse, and its change with the time delay between the two identical attosecond pulses presents an interferometric autocorrelation (IAC) curve as shown in Fig.~\ref{figure1}. There is a slight difference between the results of the TDSE and TDPT at $\tau=100\,\rm{as}$, which is due to the incompleteness of the intermediate states and the underestimation of electron correlation in the TDPT in the case of shorter pulses, but the difference is negligible at $\tau=160\,\rm{as}$. Moreover, due to the constraint imposed by the frequency limit of $\omega>I_{p2}$, the IAC curves of photoelectrons generated by TPDI induced by the attosecond pulses with a center frequency ranging from $I_{p}/2$ to $I_{p2}$ cannot be obtained through TDPT. The red dashed line represents the upper-envelope of the IAC curve of photoelectrons generated by TPDI, and its FWHM is defined as FWHM of the IAC curve of photoelectrons, which is denoted by $\tau_{e}$. If the relationship between  $\tau_{e}$ and $\tau$ is known, then the FWHM of the attosecond pulse to be measured can be obtained. 

\begin{figure}[!tb]
\includegraphics[width=75mm]{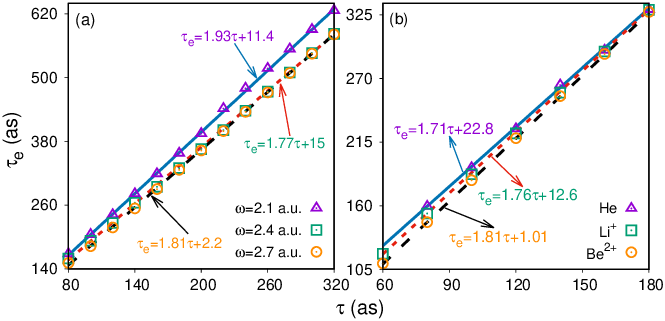}
\caption{\label{figure2} The relationship between the FWHM $\tau_{e}$ of the IAC curves of photoelectrons generated by TPDI and the FWHM $\tau$  of the attosecond pulses is approximately a straight line in the following cases: (a) different center frequencies; (b) different target ions. The peak intensity and CEP of the all attoseconds pulses are $1\times10^{14}\,\rm{Wcm^{-2}}$ and 0, respectively. For cases of helium atoms, $\rm Li^{+}$ and $\rm Be^{2+}$ ions, the corresponding center frequencies of the attosecond pulses are 2.4 a.u., 5 a.u. and 9.5 a.u., respectively. }
\end{figure} 

By fitting the data of $\tau_{e}$ with different center frequencies $\omega$ and FWHMs $\tau$ of attosecond pulses, it is found that the relationship between $\tau_{e}$ and $\tau$ can be approximately presented as a stright line as shown in Fig.~\ref{figure2} (a). When the center frequency of the attosecond pulse is greater than the second ionization potential $I_{p2}$ of the helium atom, the lines representing the relationship between $\tau_{e}$ and $\tau$ at center frequencies $\omega=2.4\,\rm a.u.$ and $2.7\,\rm a.u.$ are roughly coincident as shown by the red dotted line and the black dashed line in Fig.~\ref{figure2}(a), and the  equations for the lines are $\tau_{e}=1.77\tau+15$ and $\tau_{e}=1.81\tau+2.2$. However, in the case of longer attosecond pulse, if the center frequency of the attosecond pulse is slightly greater than $I_{p2}$, such as $\omega=2.1\,\rm a.u.$, then the relationship between $\tau_{e}$ and $\tau$ is evidently different from the cases of $\omega=2.4\,\rm a.u.$ and $\omega=2.7\,\rm a.u.$. The reason for this difference is that the value of $\sigma^{\rm{He^{+}}}(E)$ close to the ionization threshold is larger than the value of $\sigma^{\rm{He^{+}}}(E)$ far from the ionization threshold, and the increase of $\sigma^{\rm{He^{+}}}(E)$ close to the ionization threshold  increases the TPDI probability according to Eq. (\ref{equation4}), which  further leads to an increase in $\tau_{e}$ and the difference between $\tau_{e}$ and $\tau$. Shorter attosecond pulses with a center frequency close to the ionization threshold have a larger photoionization cross section $\sigma^{\rm{He^{+}}}(E)$ but a narrower effective spectrum because frequency components with energies lower than $I_{p2}$ are not able to excite the TPDI process according to TDPT,  while all the frequency components in attosecond pulses with a center frequency far away from the ionization threshold are able to excite TPDI but have smaller photoionization cross sections. Since the effect that the photoionization cross section increases offsets the effect that the effective spectral width of the attosecond pulses decrease as the photon energy gets closer to the ionization threshold, the straight lines of the relationship between $\tau_{e}$ and $\tau$ at the frequencies $\omega=2.1\,\rm a.u.$, $2.4\,\rm a.u.$ and $2.7\,\rm a.u.$ tend to coincide as $\tau$ gradually decreases as shown in Fig.~\ref{figure2} (a). On the other hand, because $\tau_{e}$ calculated by TDSE is slightly greater than that calculated by TDPI in the case of $\tau=100\,\rm{as}$ and $\omega=2.4\,\rm a.u.$ as shown in Fig.~\ref{figure1}(a), the actual $\tau$ is slightly greater than the predicted values of the equation $\tau_{e}=1.77\tau+15$ for the case of shorter attsecond pulses. Because the TDPT calculation underestimates the electron correlation for shorter attosecond pulses, the calculated $\tau_{e}$ is usually larger than the real one. The discrepancy between the calculated and real values can be considered as an indication of the strength of the electron correlation. Fig.~\ref{figure2}(b) shows the relationship between $\tau_{e}$ and $\tau$ with different target ions. It shows that the lines of different ions tend to coincide as $\tau$ gradually increases, because the spectral width of attosecond pulses decreases with increasing pulse width, resulting in a more uniform and higer TPDI probability according to Eq. (\ref{equation4}).

\begin{figure}[!tb]
\includegraphics[width=85mm]{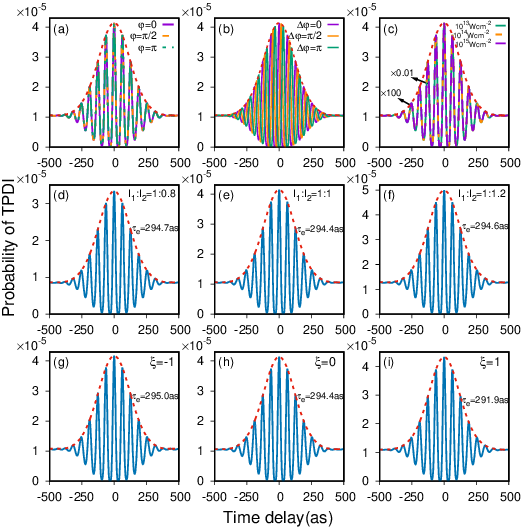}
\caption{\label{figure3} The IAC curves of photoelectrons generated by TPDI induced by attosecond pulses with different parameters: (a) different CEPs; (b)two attosecond pulses have  different CEPs; (c) different peak intensity; (d)-(f)different ratios of peak intensity; (g)-(i) different chirp parameters. The other parameters of attosecond pulses are the same as Fig.~\ref{figure1}(b).}
\end{figure}

It should be emphasized that the duration of attosecond pulses with center frequency less than the $I_{p}/2$ can be measured using the IAC curves of photoelectrons generated by TPDI through changing the target gas. Because the occurrence of TPDI relies on the energy conservation relationship $E_{1}+E_{2}=2\omega-I_{p}$, the IAC  measurement of photoelectrons induced by attosecond pulses with lower center frequency should choose target gas with lower ionization potential, such as argon, krypton, or xenon. In addition,  the attosecond pulses with ultrabroad spectral bandwidth could lead to one-photon double ionization in the IAC measurement experiment, therefore the signal of one-photon double ionization should be eliminated from the experiment data.

The IAC curves and its FWHMs of photoelectrons induced by attosecond pulses with different parameters are shown in Fig.~\ref{figure3}. Fig.~\ref{figure3}(a) shows that the IAC curve does not change with the CEP of the attoseceond pulse. However, when the CEP difference of the two time-delayed attosecond pulses changes, the peaks of the IAC curve are shifted within the same envelope as shown in Fig.~\ref{figure3}(b). Fig.~\ref{figure3}(c) presents that the amplitude of the IAC curve is proportional to the square of the attosecond pulse intensity. When the intensity ratio of the two attosecond pules decreases, the amplitude of the IAC curve increases as shown in Fig.~\ref{figure3}(d-f), but its FWHM remains virtually unchanged. It should  be noted that the FWHM of the IAC curve decreases when the chirp parameter of the attosecond pulse increases as shown in Fig.~\ref{figure3}(g-i). Therefore, the relationship $\tau_{e}=1.77\tau+15$ stays the same for the change of CEP, CEP difference, intensity, and intensity ratio of the attosecond pulses, but  changes as the chirp of the attosecond pulses changes.

\begin{figure}[!tb]
\includegraphics[width=75mm]{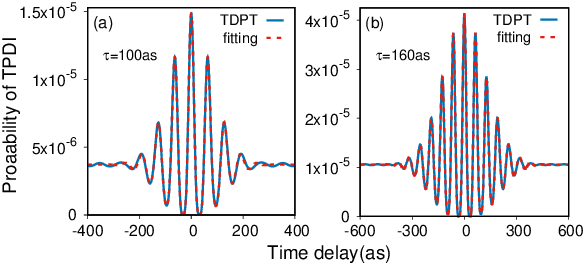}
\caption{\label{figure4} The IAC curves of photoelectrons generated by TPDI. The bule solid line and red dashed line represent the results of TDPT and fitting, respectively. The parameters of attosecond pulses are the same as Fig.~\ref{figure1}.}
\end{figure}

\begin{table}[!b]
\caption{Comparison of attosecond pulse parameters of original input and obtained by autocorrelation measurement. The $\omega$ and $\tau$ represent the center frequency and FWHM of the attosecond pulse of original input respectively, the $\omega_{m}$ and $\tau_{m}$ represent the center frequency and FWHM of the attosecond pulse obtained from autocorrelation measurement respectively, the $\eta_{\omega}$ and $\eta_{\tau}$ show the relative errors of center frequency and FWHM respectively.}
\label{table1}
\begin{ruledtabular}
\begin{tabular}{cccccc}
\textrm{$\omega$(a.u.)}&
\textrm{$\tau$(as)}&
\textrm{$\omega_{m}$(a.u.)}&
\textrm{$\tau_{m}$(as)}&
\textrm{$\eta_{\omega}$}&
\textrm{$\eta_{\tau}$}\\
\colrule

2.1   &   80    &   2.11   &  83.7     &0.48\%   &    4.62\%    \\
2.1   &   160   &   2.11   &  167.4    &0.48\%   &    4.62\%    \\
2.4   &   80    &   2.35   &  82.2     &2.08\%   &    2.80\%    \\
2.4   &   100   &   2.36   &  102.6    &1.67\%   &    2.56\%    \\
2.4   &   120   &   2.37   &  122.4    &1.25\%   &    2.00\%    \\
2.4   &   140   &   2.38   &  142.2    &0.83\%   &    1.59\%    \\
2.4   &   160   &   2.38   &  162.0    &0.83\%   &    1.29\%    \\
2.4   &   200   &   2.38   &  201.2    &0.83\%   &    0.63\%    \\
2.7   &   80    &   2.63   &  80.5     &2.59\%   &    0.69\%    \\
2.7   &   160   &   2.68   &  159.9    &0.74\%   &    0.07\%    \\
\end{tabular}
\end{ruledtabular}
\end{table}

\begin{figure}[!htb]
\includegraphics[width=75mm]{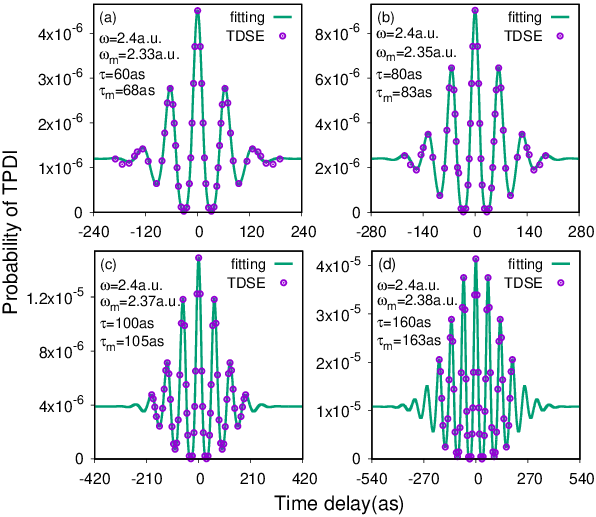}
\caption{\label{figure5} The IAC of photoelectrons generated by TPDI. The purple circles and green solid line represent the results of TDSE and fitting, respectively. The peak intensity, center frequency and CEP of the attosecond pulses are $1\times10^{14}\,\rm{Wcm^{-2}}$, 2.4 a.u. and 0, respectively. The FWHM of the attosecond pulses are: (a) 60 as; (b) 80 as; (c) 100 as; (d) 160 as. The meanings represented by $\omega$, $\tau$, $\omega_{m}$ and $\tau_{m}$ are the same as those in Table \ref{table1}.}
\end{figure}

Next, we will study the direct extraction of attosecond pulse information from IAC curves of photoelectrons. The IAC curve of a Gaussian laser pulse can be written as
\begin{equation*}
\label{equation60}
\begin{aligned}
I_{F}(t_{d})&=\int_{-\infty}^{\infty}\big|[E(t)+E(t-t_{d})]^{2}\big|^{2}dt\\
\end{aligned}
\end{equation*}
\begin{equation}
\label{equation6}
\begin{aligned}
&\hspace{0mm}=\frac{3\sqrt{\pi}E_{0}^{4}\tau}{8\sqrt{2\ln 2}}\big[1+e^{-(2\ln 2)t_{d}^{2}/\tau^{2}}+2e^{-(2\ln 2)t_{d}^{2}/\tau^{2}}\times\\
      &\hspace{5mm}\cos^{2}\omega t_{d}+4e^{-(3/4)(2\ln 2)t_{d}^{2}/\tau^{2}}\cos\omega t_{d}\big],
\end{aligned}
\end{equation}
where $E(t)=E_{0}e^{-(2\ln 2)t^{2}/\tau^{2}}\cos(\omega t+\varphi)$ is the electric field of the Gaussian laser pulse, and $t_{d}$ is the time delay. Inspired by Eq. (\ref{equation6}), a fitting function for the IAC of photoelectrons generated by TPDI can be expressed as
\begin{equation}
\label{equation7}
\begin{aligned}
f(t_{d})&=c_{1}+c_{2}e^{-(2\ln 2)t_{d}^{2}/\tau^{2}}+c_{3}e^{-(2\ln 2)t_{d}^{2}/\tau^{2}}\times\\
      &\hspace{5mm}\cos^{2}\omega t_{d}+c_{4}e^{-(2\ln 2)d_{1}t_{d}^{2}/\tau^{2}}\cos\omega t_{d}.
\end{aligned}
\end{equation}
Then the IAC curves of photoelectrons simulated by the TDPT are fitted by the fitting function of Eq. (\ref{equation7}), and the fitting results are excellent as shown in Fig.\ref{figure4}. It is found that the parameters $\omega$ and $\tau$ in the fitting function may indicate the center frequency and the FWHM of the attosecond pulse, respectively. In order to confirm the physical meaning of $\omega$ and $\tau$ in the fitting function, the IAC curves of photoelectrons induced by other attosecond pulses are fitted, and the results are shown in Table \ref{table1}. It can be seen that the parameters $\omega$ and $\tau$ are really close to the center frequency and the FWHM of the attosecond pulse, and the relative error of the fitting results  increases with the decrease of the FWHM or the center frequency as shown in Table \ref{table1}, which is due to the underestimation of the electron correlation in the TDPT. Because the experiment measurement results include the complete electron correlation, the relative error decreases for longer pulses and higher center frequencies. On the other hand, the IAC data of photoelectrons simulated by the TDSE can be also fitted by Eq. (\ref{equation7}), and the fitting results are good as shown in Fig.\ref{figure5}. These results indicate that IAC measurements based on TPDI  can characterize attosecond pulses with FWHM ranging from tens of attosecond to hundreds of attosecond.

The IAC curve of a chirped Gaussian laser pulse can be writen as
\begin{equation}
\label{equation8}
\begin{aligned}
I_{F}(t_{d})&=\int_{-\infty}^{\infty}\big|[E(t)+E(t-t_{d})]^{2}\big|^{2}dt\\
      &=\frac{3\sqrt{\pi}E_{0}^{4}\tau}{8\sqrt{2(1+{\xi}^2)\ln 2}}\Bigg\{1-e^{-\frac{(2\ln 2)t_{d}^{2}}{\tau^{2}}}+2e^{-\frac{(2\ln 2)t_{d}^{2}}{(1+{\xi}^2)\tau^{2}}}+\\
      &\hspace{5mm}2e^{-\frac{(2\ln 2)t_{d}^{2}}{\tau^{2}}}\cos^{2}\omega t_{d}+2e^{-\frac{(2\ln 2)(3+{\xi}^2)t_{d}^{2}}{4(1+{\xi}^2)\tau^{2}}}\times\\
      &\hspace{5mm}\cos\Bigg[\omega t_{d}-\frac{{\xi}t_{d}^{2}\ln 2}{(1+{\xi}^2)\tau^{2}}\Bigg]+2e^{-\frac{(2\ln 2)(3+{\xi}^2)t_{d}^{2}}{4(1+{\xi}^2)\tau^{2}}}\times\\
      &\hspace{5mm}\cos\Bigg[\omega t_{d}+\frac{{\xi}t_{d}^{2}\ln 2}{(1+{\xi}^2)\tau^{2}}\Bigg]\Bigg\},
\end{aligned}
\end{equation}
where $E(t)=\frac{E_{0}}{\sqrt[4]{1+{\xi}^2}}e^{-\frac{(2\ln 2)t_{d}^{2}}{(1+{\xi}^2)\tau^{2}}}\cos\Big\{\Big[\omega+\frac{(2\ln 2)\xi}{(1+{\xi}^2)\tau^{2}}\Big]t+\varphi\Big\}$\cite{peng2009few} is the electric field of the chirped Gaussian laser pulse, and $t_{d}$ is the time delay. Similarly, a fitting function for the IAC of photoelectrons generated by TPDI can be expressed as
\begin{equation}
\label{equation9}
\begin{aligned}
f(t_{d})&=a_{1}+a_{2}e^{-\frac{(2\ln 2)t_{d}^{2}}{\tau^{2}}}+a_{3}e^{-\frac{(2\ln 2)d_{1}t_{d}^{2}}{(1+{\xi}^2)\tau^{2}}}+\\
      &\hspace{5mm}a_{4}e^{-\frac{(2\ln 2)t_{d}^{2}}{\tau^{2}}}\cos^{2}\omega t_{d}+a_{5}e^{-\frac{(2\ln 2)d_{2}(3+{\xi}^2)t_{d}^{2}}{4(1+{\xi}^2)\tau^{2}}}\times\\
      &\hspace{5mm}\cos\Bigg[\omega t_{d}-\frac{{\xi}t_{d}^{2}\ln 2}{(1+{\xi}^2)\tau^{2}}\Bigg]+a_{5}e^{-\frac{(2\ln 2)d_{2}(3+{\xi}^2)t_{d}^{2}}{4(1+{\xi}^2)\tau^{2}}}\times\\
      &\hspace{5mm}\cos\Bigg[\omega t_{d}+\frac{{\xi}t_{d}^{2}\ln 2}{(1+{\xi}^2)\tau^{2}}\Bigg].
\end{aligned}
\end{equation}
The IAC curves of photoelectrons in Fig. \ref{figure3} (g) and (i) can be fitted by the fitting function of Eq. (\ref{equation9}), and the center frequency $\omega$ and the FWHM $\tau$ can be extracted.  However, the chirp parameters cannot be extracted.

\begin{figure}[!htb]
\includegraphics[width=85mm]{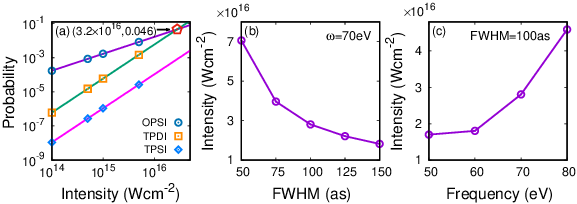}
\caption{\label{figure6} The required peak intensity of the attosecond pulse in the IAC experiments. (a) The probabilities of OPSI and TPDI as functions of the peak intensity of the attosecond pulses, where the FWHM $\tau=100\,\rm{as}$ and the center frequency $\omega=70\,\rm{eV}$, and the cross point marked by a red triangle shows the required peak intensity in the IAC experiment. (b)-(c) The required peak intensity of the autocorrelation experiment changes as the FWHM or center frequency of the attosecond pulse changes.}
\end{figure}

Finally, a rough estimate of required peak intensity of the attosecond pulse in IAC experiments is given. Because the advance of attosecond technology has allowed one to study the electron correlation and the time delay in the OPSI \cite{schultze2010delay,ossiander2017attosecond}, the signal of OPSI  can be used as a reference for the required peak intensity of the attosecond pulse in IAC experiments. Although the cross section of TPDI is rather low, high photon flux attosecond pulse can enhance the probability of TPDI. The peak intensity at the cross point of the  probability curves of OPSI and TPDI is chosen as an estimate as shown in Fig. \ref{figure6}(a), where the required peak intensity in IAC experiments is $3\times 10^{16}\,\rm{Wcm^{-2}}$ for attosecond pulses with  a FWHM $\tau=100\,\rm{as}$ and a center frequency $\omega=70\,\rm{eV}$. It is also found that the required peak intensity increases as  the FWHM decreases or the center frequency increases as shown in Fig. \ref{figure5} (b-c). The values shown in Fig.6 are on the same order as an estimate given in Ref\cite{gao2022quantification}.

\section{Conclusion}

In conclusion, autocorrelation measurement is theoretically demonstrated to be used to characterize the FWHM and the center frequency of  attosecond pulses, and the required lower limit of the peak intensity of the attosecond pulse is estimated to be on the order of $ 10^{16}\,\rm{Wcm^{-2}}$ in autocorrelation experiments. We anticipate that this work, combined with advance in high-flux attosecond pulses generation\cite{wu2013generation,li2019double,zhong2016noncollinear,chopineau2021spatio}, will stimulate researchers to carry out further experiment with the aim of verifying the availability of characterizing attosecond pulse and exploring the property of matter based on autocorrelation measurement of attosecond pulse.

\begin{acknowledgments}

This work was supported by the Synergetic Extreme Condition User Facility (SECUF), National Natural Science Foundation of China under Grant Nos.  92150103, 12204526, 12074418, 11774411, 92250303), Chinese Academy of Sciences (Grant No. YSBR-059), Ministry of Science and Technology of China (Grant No. 2022YFB4601100), and STU Scientific Research Initiation Grant(Nos. NTF23011, NTF22026, NTF23014, NTF23036T, NTF24005T).

\end{acknowledgments}



\nocite{*}

\bibliography{apssamp}

\end{document}